\documentclass[a4paper,11pt]{article}
\usepackage{pos}

\usepackage{siunitx}
\usepackage{graphicx}
\usepackage{subcaption}
\usepackage{booktabs} 
\usepackage{array} 

\title{Development of a Hybrid Scintillator Strip–Drift Tube Muon Detector for the FCC-ee}

\author[a]{Jiajin Ge}
\onbehalf{\\on behalf of the University of Michigan FCC-ee R\&D team}

\affiliation[a]{Department of Physics, University of Michigan,\\Ann Arbor, MI, United States of America}

\emailAdd{jiajinge@umich.edu}

\abstract{The Future Circular Collider (FCC-ee) is a proposed high-luminosity $e^+e^-$ collider for precision measurements at the electroweak, Higgs, and top-quark energy scales. A high-performance, cost-effective muon system is essential for muon identification, triggering, momentum measurement, and long-lived particle detection.
We present the R\&D of a hybrid muon detector concept combining fast extruded plastic scintillator strips with precision drift tubes. The scintillator strips were produced at Fermilab, while detector prototypes with SiPM readout and pressurized drift tubes were designed, constructed, and integrated at the University of Michigan. The hybrid detector was tested at the CERN test beam facility in 2025 and 2026. The scintillator strips achieved a position resolution of 1.6~mm and a time resolution of approximately 350~ps, while the drift tubes provided a single-hit spatial resolution better than 100~$\mu$m. These results demonstrate the feasibility of a high-performance hybrid muon spectrometer for the FCC-ee.}

\renewcommand{\logo}{\relax}
\begin{document}
\maketitle

\section{Introduction}

The Future Circular Collider (FCC-ee) is a proposed high-luminosity circular \(e^+e^-\) collider designed to operate at center-of-mass energies corresponding to the production of the \(Z\) boson (\(\sim91\)~GeV), \(WW\) pairs (\(\sim160\)~GeV), \(ZH\) (\(\sim250\)~GeV), and \(t\bar{t}\) (\(\sim365\)~GeV). To fully exploit its precision physics program at the electroweak, Higgs, and top-quark energy scales, the FCC-ee detector concepts under development all require advanced tracking, calorimetry, and high-performance muon systems~\cite{DAM2025170648}.

\begin{figure}[!htbp]
    \centering
    \includegraphics[width=0.75\textwidth]{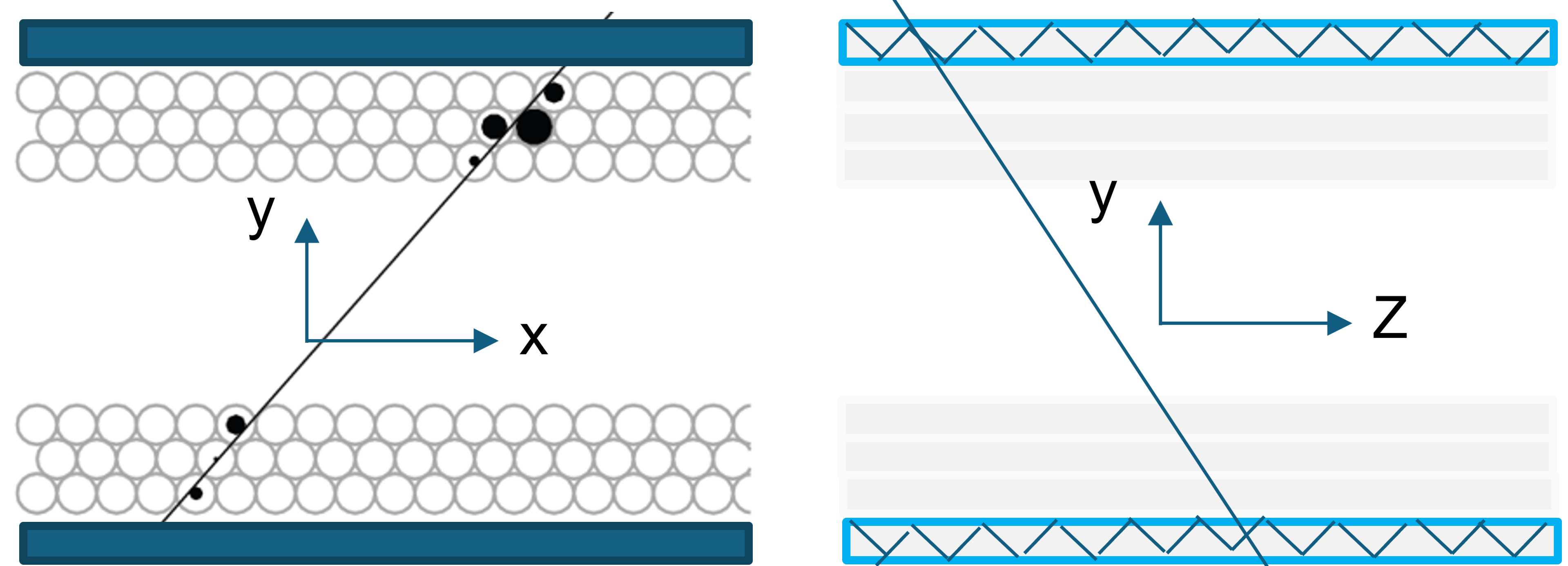}
    \caption{Schematic of the hybrid muon detector concept, consisting of six layers of precision drift tubes and orthogonal scintillator-strip layers on the top and bottom, with an example reconstructed muon track. Left: drift-tube cross section. Right: drift-tube side profile and triangular scintillator-strip cross section.}
    \label{fig:muon_detector}
\end{figure}

To develop a high-performance, cost-effective, and robust muon system for the FCC-ee, the University of Michigan (UM) FCC-ee R\&D team has proposed the hybrid detector concept shown in Figure~\ref{fig:muon_detector},
that combines fast extruded plastic scintillator strips with precision drift tubes~\cite{Anulli:2025uxj}. As the outermost detector subsystem, the muon detector provides efficient muon identification and triggering while serving as a hadron filter. Its precision tracking complements the inner tracker for momentum measurement and is particularly valuable for reconstructing tracks from long-lived particles (LLPs) that leave limited signatures in the inner detector. The combination of excellent spatial and timing resolution further enhances sensitivity to physics beyond the Standard Model by enabling time-of-flight identification of massive stable charged particles and the reconstruction of displaced vertices from LLP decays within the large muon detector volume.

These two complementary technologies have both been successfully demonstrated in large-scale particle physics experiments. The ATLAS Monitored Drift Tube (MDT) muon spectrometer~\cite{BAUER200117} has shown that pressurized aluminum drift tubes can achieve a single-tube spatial resolution of 80$\mu\text{m}$ with excellent long-term reliability. Meanwhile, the D0 Central Preshower detector~\cite{d0_scint, PLADALMAU2001482} demonstrated the effectiveness of extruded plastic scintillator strips for fast timing. Although its readout relied on cryogenic Visible Light Photon Counters (VLPCs), the recent development of silicon photomultipliers (SiPMs) provides a compact, robust, and cost-effective readout solution, making scintillator-strip detectors particularly attractive for the FCC-ee muon system.

Building upon these established technologies, our R\&D goal is to develop a high-performance muon system with precise timing resolution ($\sigma_t \sim 200,\text{ps}$) and spatial resolution better than 100$\mu\text{m}$. We will optimize the detector geometry and readout configuration to minimize channel count while preserving the required tracking and timing performance.


\section{The Proposed Muon Detector Configuration and Layout}
\label{sec:detector_concept}

As illustrated in Figure~\ref{fig:muon_detector}, each detector module consists of two drift-tube multilayers separated by a central spacer, with fast-timing layers of orthogonal scintillator strips mounted above and below. This hybrid design separates precision tracking and timing into complementary subsystems: the drift tubes provide high-resolution tracking in the transverse ($x$--$y$) plane for momentum measurement and displaced-vertex searches, while the scintillator strips measure the longitudinal ($z$) coordinate and provide precise timing, and enable low-latency triggering and cosmic-ray rejection.

\begin{figure}[htbp]
    \centering
    \includegraphics[width=0.75\textwidth]{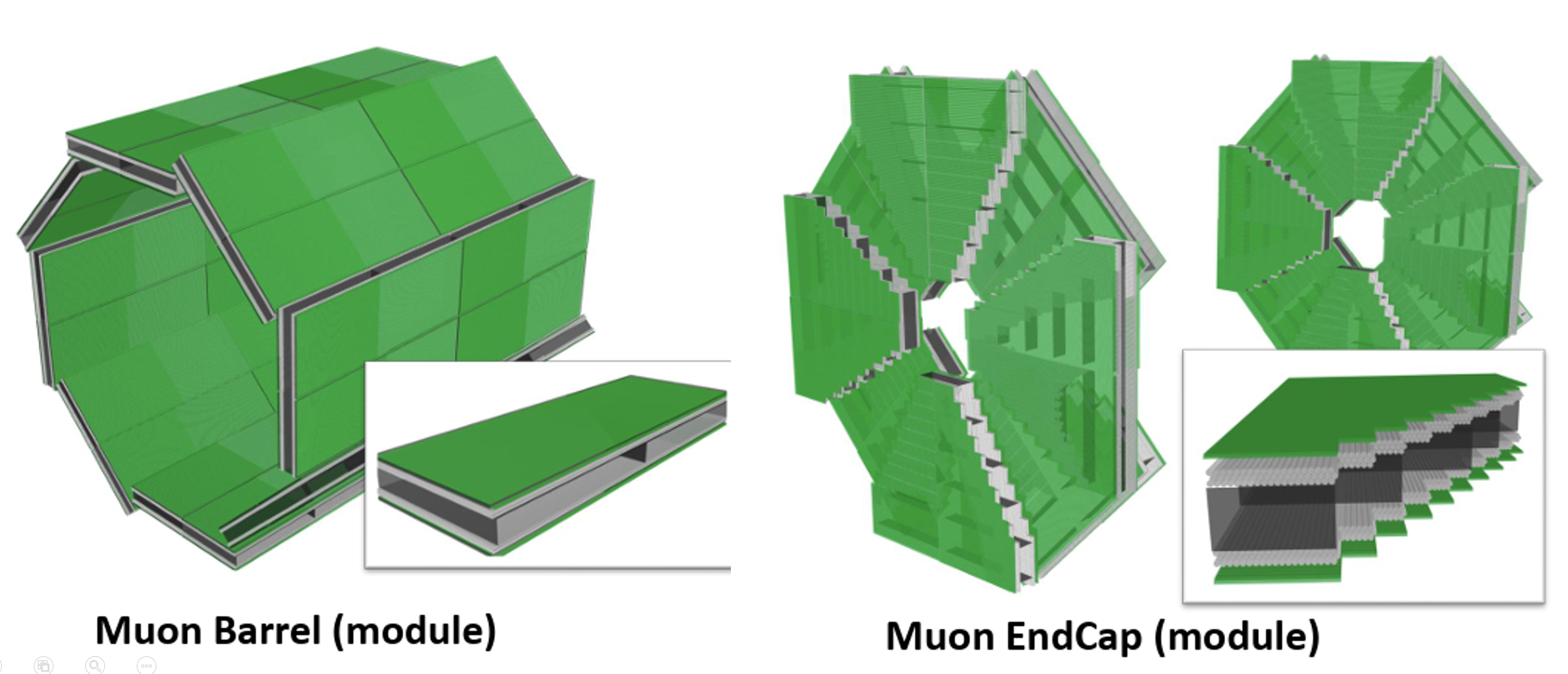}
    \caption{GEANT4-based DD4hep implementation of the ALLEGRO muon detector geometry. Left: barrel module with 30mm diameter drift tubes of 6000~mm length arranged in six layers and triangular scintillator strips of 2200~mm length; right: endcap module with drift tubes (1000--3700~mm length) and scintillator strips (586--2170~mm length) oriented perpendicular to the beam line.}
\label{fig:muon_system}
\end{figure}

To evaluate this concept at the experiment scale, the muon system has been implemented in the ALLEGRO detector framework using GEANT4-based DD4hep simulations~\cite{DD4hep:2014}. 
The detector modules are arranged in an eight-fold azimuthal geometry with overlapping coverage, providing more than 98\% geometric acceptance. As shown in Figure~\ref{fig:muon_system}, the design comprises 32 barrel modules symmetrically distributed in the $+\eta$ and $-\eta$ hemispheres and 32 modules in two endcap wheels, with 27,648 drift-tube channels and 30,080 double-ended scintillator strips, for a total of 87,808 readout channels. A single station provides efficient muon identification and calorimeter tail catching, while its timing and tracking capabilities enhance sensitivity to massive stable particles and long-lived particles. Multiple stations can be added for standalone momentum measurement when integrated with the experiment's magnet system.

\section{Scintillator Strip Detector Prototype and Readout Electronics}
\label{sec:detector_construct}

\begin{figure}[htbp]
    \centering
    \subfloat[]{
    \includegraphics[width=0.5\textwidth, height=2cm]{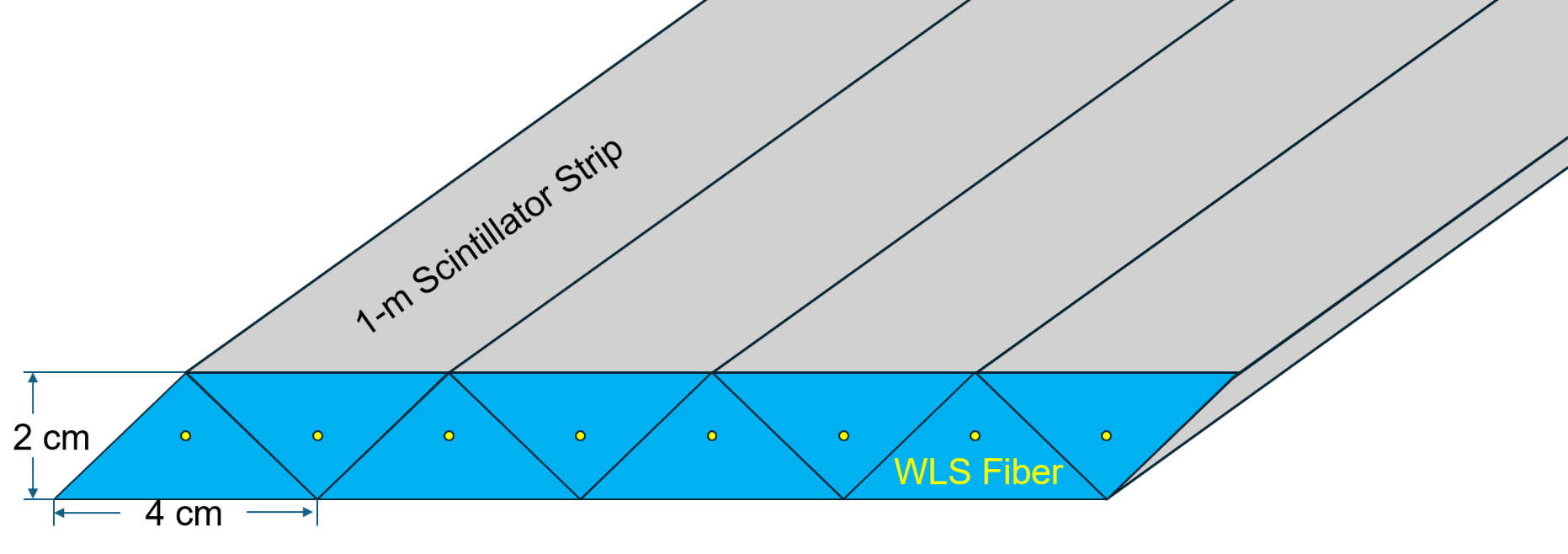} } \hspace{3mm}
   \subfloat[]{
       \includegraphics[width=0.45\textwidth, height=3cm]{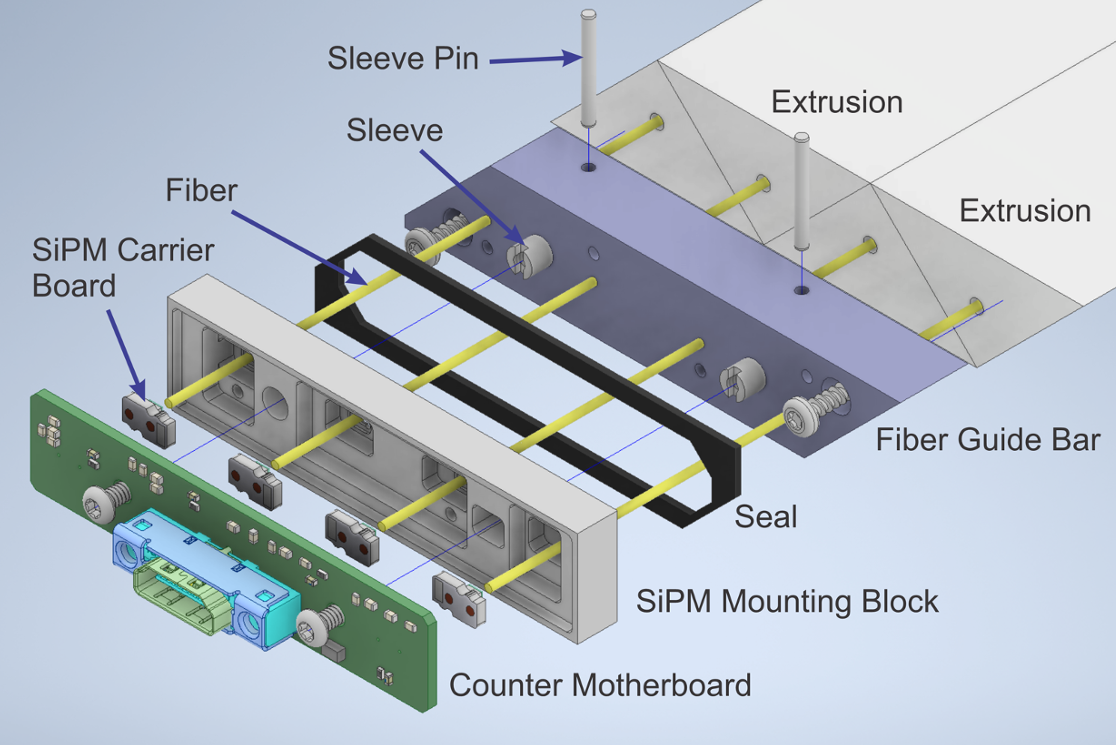}}
    \caption{(a) The 8-strip scintillator prototype module. Triangular strips (4~cm base, 2~cm height, 1~m length) are arranged in an alternating up-down overlapping configuration, with each strip housing a 1.5~mm WLS fiber in its central guide hole. (b) Exploded view of scintillator Quad-counter detector module~\cite{Bross:2022Giza}.}
    \label{fig:scintillator_strip}
\end{figure}

The fast-timing sub-system uses extruded plastic scintillator strips developed at Fermilab~\cite{PLADALMAU2001482}. Through two visits to Fermilab, we acquired 16 large triangular strips (4~cm base, 2~cm height, 1~m length) along with additional smaller strips (2~cm base, 1~cm height, 1~m length), as well as the construction expertise needed for assembly at UM. The large strips were used in the prototype, as shown in Figure~\ref{fig:scintillator_strip}(a). Each strip is wrapped in a reflective layer, $\text{TiO}_2$-coated on the top and bottom surfaces and covered with white Tyvek paper on the lateral sides, to maximize light collection and ensure optical isolation. The triangular cross-section allows adjacent strips to overlap continuously, eliminating dead zones, and exploits the vernier charge-sharing effect for precise hit reconstruction~\cite{ADAMS1996131}.
Each strip contains a central guide hole housing a 1.5~mm diameter Kuraray YS-2M or YS-4M WLS fiber~\cite{Kuraray_WLS_Fiber}. 

As shown in Figure~\ref{fig:scintillator_strip}(b),
precision-machined fiber guide bars are mounted at both ends of each 4-strip unit to secure the fibers, with end faces diamond-polished to maximize photon transmission. Custom SiPM carriers align Hamamatsu S13360-3075PE SiPMs ($3 \times 3\,\text{mm}^2$, $75\,\mu\text{m}$ pitch) with the fiber ends, using Cargille optical gel for coupling. A readout PCB motherboard at each end routes bias voltage and analog signals to the front-end electronics; the PCB also includes an LED calibration circuit with each LED illuminating four adjacent strips. Two 4-strip units are combined to form an 8-strip prototype module. 

We adopt a double-ended readout scheme for the scintillator strips: the arrival time difference at the two SiPMs yields the longitudinal impact coordinate, while the average time provides the timing measurement. The SiPM signals are transmitted via 2~m coaxial cables to the CAEN DT5202 front-end module~\cite{CAEN_DT5202}, a 64-channel unit housing two Citiroc-1A chips. Each channel features a High-Gain (HG) path for single photoelectron calibration and a Low-Gain (LG) path for Minimum Ionizing Particle (MIP) signals, with each channel comprising a preamplifier, a slow shaper with peak-sensing detector, and a fast shaper followed by a discriminator. The on-board FPGA achieves a Time-of-Arrival (TOA) and Time-over-Threshold (TOT) resolution of approximately 0.25~ns. A single DT5202 connects to a PC via Ethernet; for larger systems, multiple units can be daisy-chained through optical links.

\section{Drift Tube Detectors}

Gaseous drift tubes have been widely adopted in colliding beam experiments for their mechanical robustness, long-term operational stability, and low fabrication cost. The tracking subsystem draws upon the technological heritage of the ATLAS MDT spectrometer. The ATLAS MDT uses 30~mm aluminum drift tubes with a 3~bar $\text{Ar}:\text{CO}_2$ (93\%:7\%) gas mixture to achieve an intrinsic single-tube resolution of $80\,\mu\text{m}$, and has demonstrated exceptional reliability with a $99.7\%$ functional channel fraction after more than 18 years of operation. For the HL-LHC upgrade, smaller-diameter sMDTs (15~mm) have been developed with a resolution of $\sim 82\,\mu\text{m}$.

\begin{figure}[htbp]
    \centering
        \subfloat[]{
    \raisebox{5mm}{
        \includegraphics[width=0.4\textwidth, height=3.0cm]{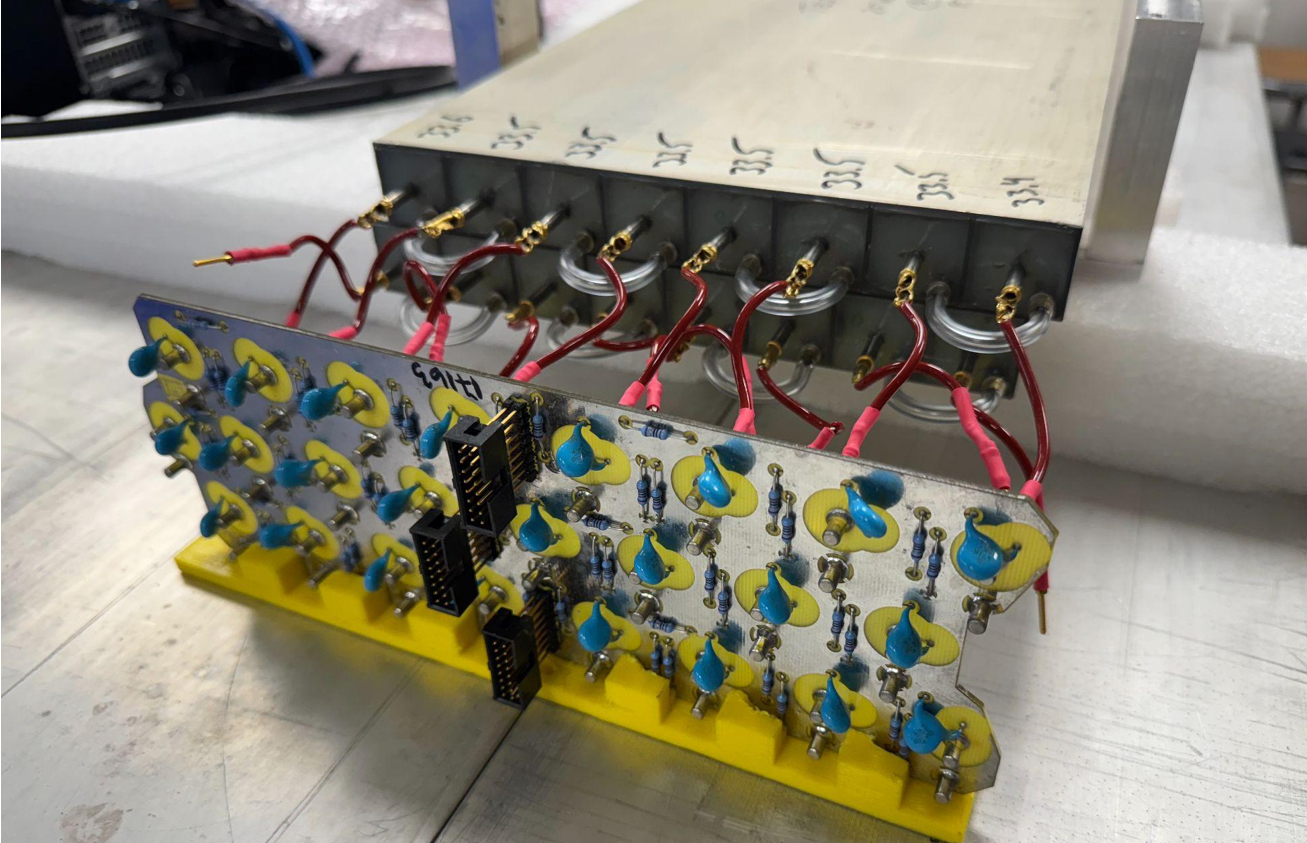}
        \label{fig:squared_tube}
    }}
    \subfloat[]{
        \includegraphics[width=0.43\textwidth,height=3.5cm]{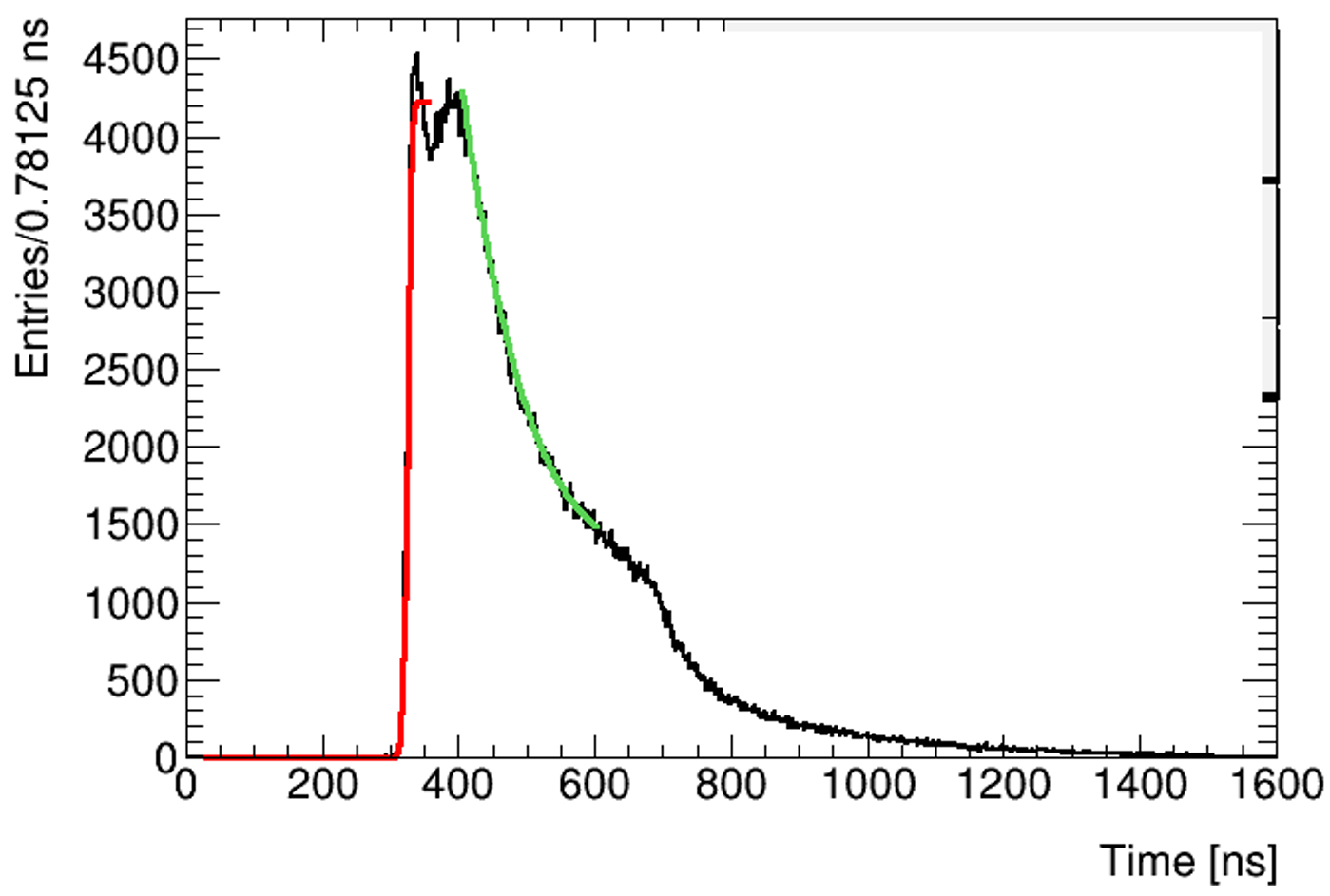}
        \label{fig:squared_tube_TDC}
    }
    
\caption{(a) Photograph of the 1-inch square drift tube detector prototype, read out by the new ATLAS MDT electronics. The prototype will be shielded by Faraday cages for noise suppression. (b) TDC spectrum measured with cosmic rays, showing a long tail in the drift time distribution. The tail is attributed to the weakened electric field near the corners of the square profile, which delays the collection of ionization electrons from these regions.}    
\label{fig:squared_tube}
\end{figure}

We are exploring two complementary approaches for a cost-effective muon detector: reusing existing ATLAS MDT chambers and developing new low-cost rectangular drift tubes. Because the momentum resolution requirements at an $e^+e^-$ Higgs factory are less stringent, 1-inch square drift tubes offer a promising alternative, simplifying chamber assembly and potentially reducing manufacturing costs by up to a factor of five while maintaining competitive performance. A square-tube prototype was built (Figure~\ref{fig:squared_tube}(a)) and tested at UM using an eight-layer sMDT tracker as a reference, providing approximately $100\mu\text{m}$ spatial resolution. At an operating pressure of 1.65~bar and a voltage of 2176~V, the tube exhibits high ionization efficiency. However, electric-field distortions near the corners complicate the drift-time response, resulting in tails in the TDC spectrum, as shown in Figure~\ref{fig:squared_tube_TDC}. Ongoing studies focus on optimizing the grounding configuration and developing calibration algorithms to mitigate these effects.

In parallel, a gas mixing system has been built at UM to study drift gas compositions and explore environmentally friendly alternatives. Using sMDT chambers at 3~bar, we observe that increasing the $\text{CO}_2$ fraction improves tracking resolution at the cost of longer drift times.

\section{Test beam campaign and preliminary results}
\begin{figure}[htbp]
    \centering
    \subfloat[]{
        \includegraphics[width=0.45\textwidth,height=4cm]{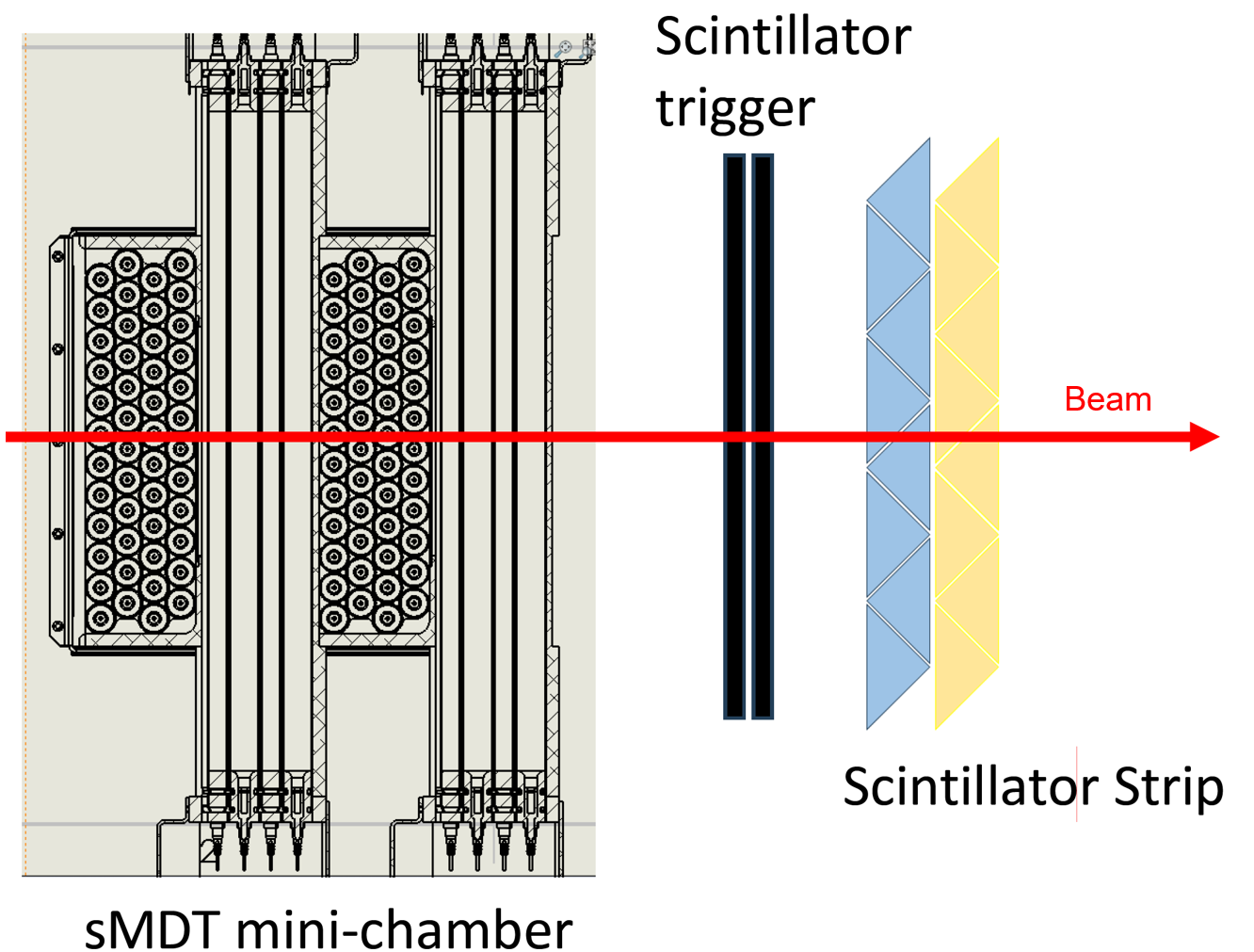}
        \label{fig:TB_Setup_Schematic}
    }\hspace{5mm}
    \subfloat[]{
        \includegraphics[width=0.32\textwidth,height=4cm]{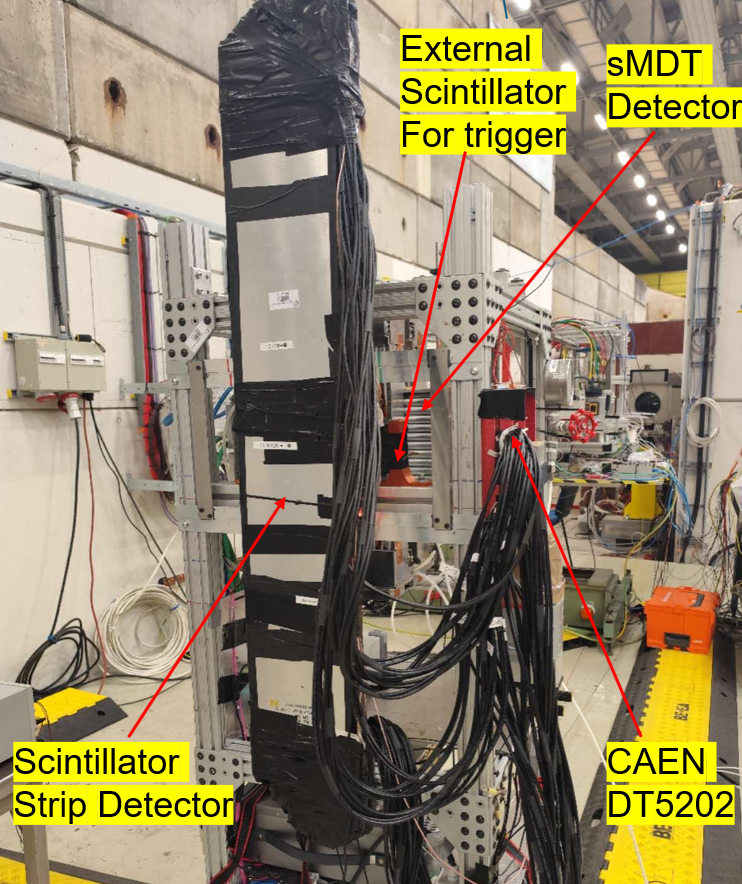}
        \label{fig:TB_Setup_Picture}
    }  
\caption{(a) Top view of the test beam setup. Four mini sMDT chambers provide 3D tracking, while two vertically oriented 8-channel scintillator-strip detectors with double-ended SiPM readout are used for timing measurements. Two external scintillators provide the coincidence trigger. (b) Photograph of the setup.}
    \label{fig:TB_Setup}
\end{figure}

The performance of the built prototype modules was tested during two test beam campaigns at the CERN Proton Synchrotron (PS) T9 facility in November 2025 and March 2026, using a 10~GeV hadron beam. The test beam setup is shown in Figure~\ref{fig:TB_Setup}. It integrated two 8-channel scintillator modules, forming a 16-channel array. Precision tracking was provided by four mini sMDT chambers (two vertical and two horizontal), operated with a 3~bar $\text{Ar}:\text{CO}_2$ (93\%:7\%) gas mixture at +2730~V. These sMDT chambers achieved an intrinsic 3D tracking resolution of approximately $85\,\mu\text{m}$, serving as an ideal spatial anchor for the beam track reconstruction. Two external plastic scintillators were used in coincidence to generate the trigger.

\begin{figure}[!htbp]
    \centering
    \subfloat[]{
        \includegraphics[width=0.2\textwidth]{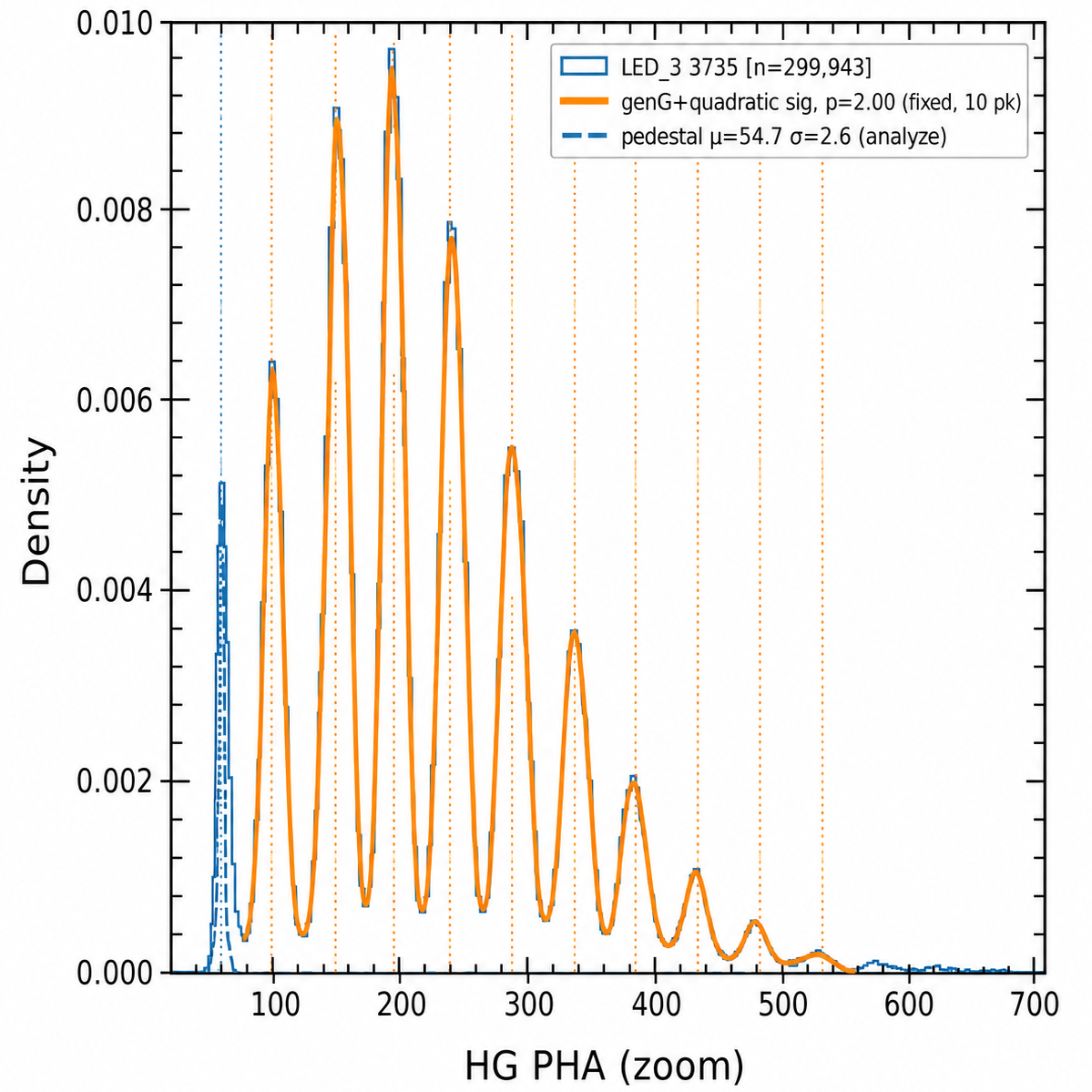}
        \label{fig:LED_calibration}
    }
    \subfloat[]{
        \includegraphics[width=0.25\textwidth]{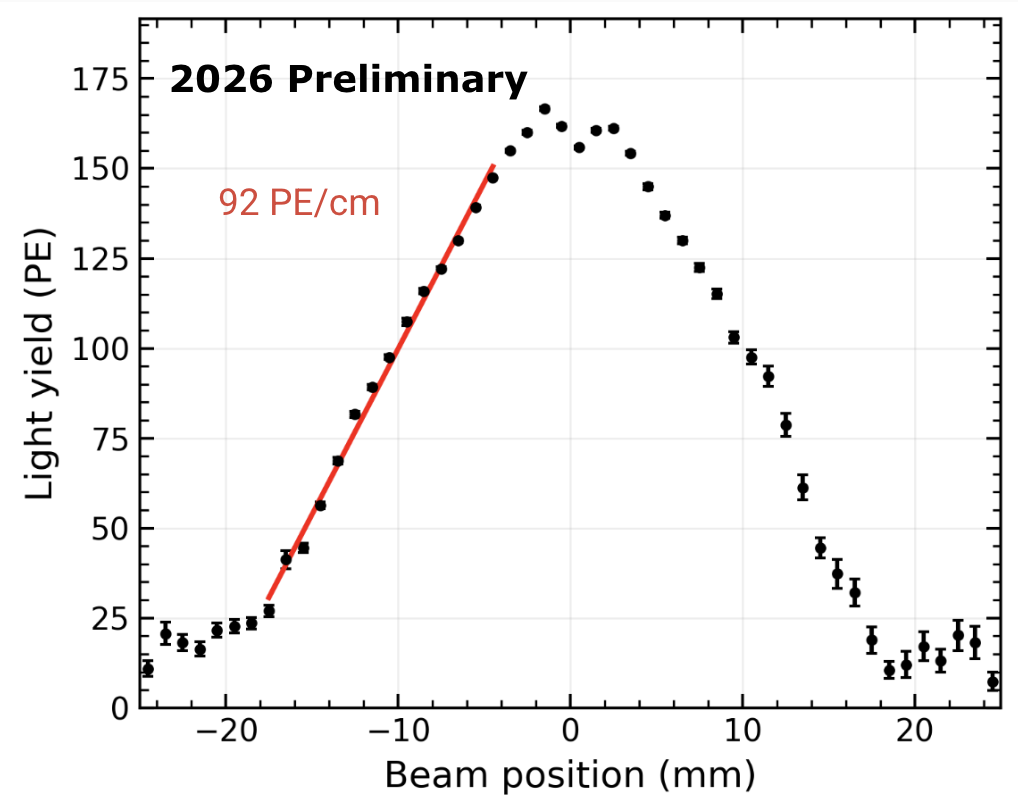}
        \label{fig:TB_PE-yield}
    }
    \subfloat[]{
        \includegraphics[width=0.25\textwidth]{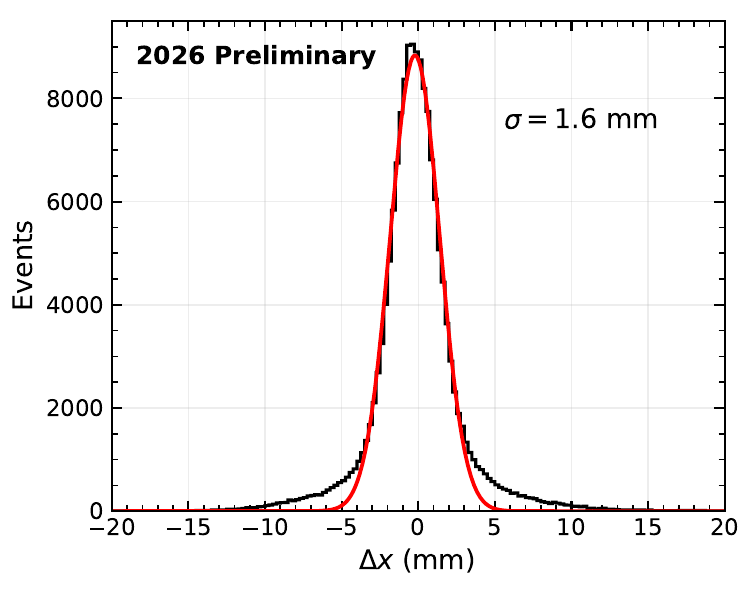}
        \label{fig:TB_posi_res}
    }
    \subfloat[]{
        \includegraphics[width=0.25\textwidth]{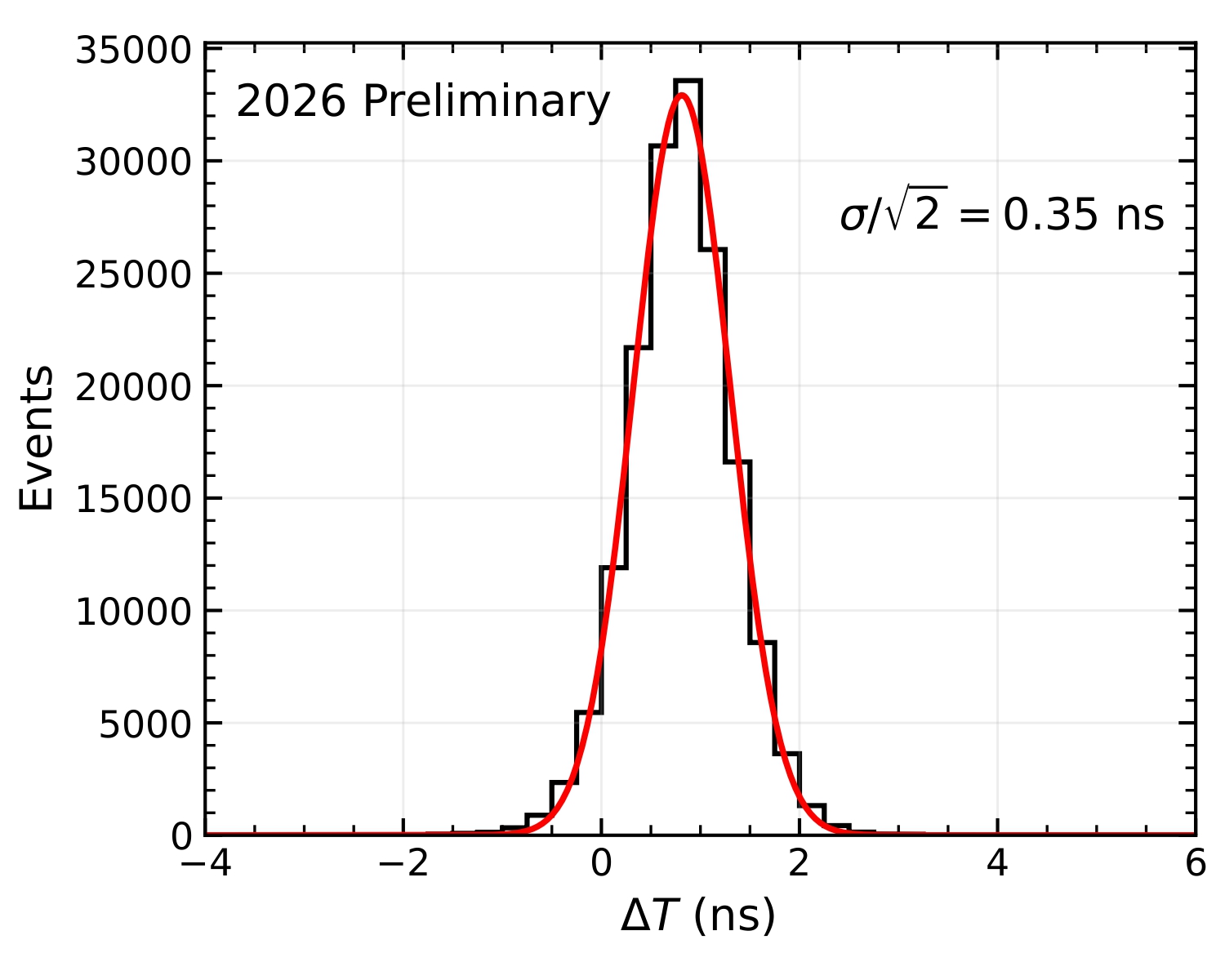}
        \label{fig:TB_time_res}
    }
    \caption{(a) High-gain (HG) LED calibration spectrum showing the single-photoelectron (PE) peaks used for channel gain equalization. (b) Light yield as a function of beam position across the triangular scintillator strip. (c) Spatial residual distribution of the scintillator strip detector relative to the sMDT reference track, for events with hits in two neighboring strips, corresponding to a spatial resolution of 1.6~mm. (d) Single-channel time resolution measured during the March~2026 test beam campaign, demonstrating a timing resolution of approximately 350~ps.}
    \label{fig:TB_results}
\end{figure}

A full spatial scan was performed by moving the detector prototype through 8 distinct longitudinal positions along the strip length and 7 transverse positions across the strip width. Angular scans were performed by tilting the detector frame up to $\pm 30^\circ$ relative to the beam axis. After every movement, an LED calibration run was executed to provide channel-by-channel gain calibration for subsequent data processing, as illustrated in Figure~\ref{fig:LED_calibration}.
The light yield as a function of beam position across the triangular scintillator strip (Figure~\ref{fig:TB_PE-yield}) shows 92 PE/cm.

The transverse impact position was reconstructed using a charge-weighted average of the deposited charges in the fired strips:
\begin{equation}
    x = \frac{\sum_i Q_i x_i}{\sum_i Q_i},
\end{equation}
where $Q_i$ is the LED-calibrated LG amplitude in strip $i$, and $x_i$ is the center position of strip $i$, with $x_i = x_0 + i \cdot (w/2)$, where $x_0$ is the position of the leftmost strip center and $w$ is the base width of the triangular strip. The spatial resolution of the scintillator strip detector was evaluated by comparing the reconstructed hit position from the scintillator with the reference track position from the sMDT chambers. Figure~\ref{fig:TB_posi_res} shows the residual distributions for the one module, combining the LG results from both ends of the strips for events with hits in two neighboring strips. A Gaussian fit yields a core resolution of approximately $1.6$~mm, demonstrating the effectiveness of the vernier charge-sharing reconstruction with LED-calibrated amplitudes. The single-channel time resolution measured during the March 2026 test beam campaign reached approximately 350~ps, as shown in Figure~\ref{fig:TB_time_res}. With full time-slew corrections, low-noise electronics, and high-precision time measurement, the timing performance is expected to approach the 200~ps design target.

\section{Conclusions}
We have developed a hybrid muon detector concept for the FCC-ee experiment, combining precision drift tubes with fast scintillator strips for highly efficient muon identification, triggering, and 3-dimensional tracking.  The preliminary design provides near-complete solid-angle coverage with a total of 87,808 readout channels. ATLAS MDT technology serves as the baseline, while 1-inch square tubes are being explored as a cost-effective alternative. Two scintillator prototypes using triangular strips and double-ended SiPM readout achieved a spatial resolution of $1.6$~mm and a preliminary time resolution of $350$~ps in test beams, demonstrating the feasibility of this hybrid approach for a cost-effective FCC-ee muon detector.

\section*{Acknowledgments}
We thank Alan Bross and Jim Freeman of Fermilab and E. Craig Dukes of the University of Virginia for their assistance with the material and design of the scintillator strip prototypes.
This work was supported in part by the U.S. Department of Energy under Grant DE-SC0007859.

\bibliographystyle{JHEP}
\bibliography{references.bib}

@article{DAM2025170648,
  title     = {{Detector requirements, design, and technologies for the FCC-ee Higgs, electroweak, and top factory}},
  journal   = {Nucl. Instrum. Meth. A},
  volume    = {1080},
  pages     = {170648},
  year      = {2025},
  issn      = {0168-9002},
  doi       = {10.1016/j.nima.2025.170648},
  author    = {Mogens Dam}
}

@article{d0_scint,
  author    = {{D0 Collaboration}},
  title     = {{The Upgraded D0 Detector}},
  journal   = {Nucl. Instrum. Meth. A},
  volume    = {565},
  pages     = {463--537},
  year      = {2006},
  doi       = {10.1016/j.nima.2006.05.248}
}

@article{PLADALMAU2001482,
title = {Low-cost extruded plastic scintillator},
journal = {Nucl. Instrum. Meth. A},
volume = {466},
number = {3},
pages = {482-491},
year = {2001},
issn = {0168-9002},
doi = {https://doi.org/10.1016/S0168-9002(01)00177-2},
author = {Anna Pla-Dalmau and others}
}

@article{BAUER200117,
title = {Construction and test of MDT chambers for the ATLAS muon spectrometer},
journal = {Nucl. Instrum. Meth. A},
volume = {461},
number = {1},
pages = {17-20},
year = {2001},
note = {8th Pisa Meeting on Advanced Detectors},
issn = {0168-9002},
doi = {https://doi.org/10.1016/S0168-9002(00)01156-6},
author = {F. Bauer and others}

}

@article{DD4hep:2014,
  author  = "Frank, M. and others",
  title   = "{DD4hep: A Detector Description Tool-kit for High Energy Physics Experiments}",
  journal = "J. Phys. Conf. Ser.",
  volume  = "513",
  pages   = "022010",
  year    = "2014",
  doi     = "10.1088/1742-6596/513/2/022010"
}

@article{Kuraray_WLS_Fiber,
    author = {Kodama, Shoma and others},
    title = {Performance of New Kuraray Wavelength-shifting Fibers with Short Decay Time},
    journal = {Prog. Theor. Exp. Phys.},
    volume = {2024},
    number = {5},
    pages = {053H01},
    year = {2024},
    month = {05},
    issn = {2050-3911},
    doi = {10.1093/ptep/ptae055}
}

@article{ADAMS1996131,
title = {A new detector technique using triangular scintillating strips to measure the position of minimum ionizing particles},
journal = {Nucl. Instrum. Meth. A},
volume = {378},
number = {1},
pages = {131-142},
year = {1996},
issn = {0168-9002},
doi = {https://doi.org/10.1016/0168-9002(96)00425-1},
author = {M Adams and others}
}

@misc{CAEN_DT5202,
    author = {{CAEN Technologies}},
    title = {DT5202 - Desktop 64 Channel Citiroc unit for FERS-5200},
    year = {2026},
    url = {https://caen.it/products/dt5202/},
    note = {Accessed: 2026-07-21}
}

@article{Anulli:2025uxj,
    author = "Anulli, F. and others",
    title = "{A High-Precision, Fast, Robust, and Cost-Effective Muon Detector Concept for the FCC-ee}",
    eprint = "arXiv:2504.10448",
    archivePrefix = "arXiv",
    primaryClass = "hep-ex",
    reportNumber = "FERMILAB-PUB-25-0248-PPD",
    month = "4",
    year = "2025"
}

@article{Bross:2022Giza,
  author = {Bross, Alan D. and others},
  title = {Tomographic Muon Imaging of the Great Pyramid of Giza},
  journal = {Journal of Advanced Instrumentation in Science},
  volume = {2022},
  pages = {280},
  year = {2022},
  doi = {10.31526/jais.2022.280},
  eprint = {2202.08184},
  archivePrefix = {arXiv},
  primaryClass = {physics.ins-det}
}

\end{document}